\documentclass[aps,showpacs,superscriptaddress, twocolumn]{revtex4}
\usepackage{graphicx}
\usepackage{amsmath}
\usepackage{amssymb}
\usepackage{tabularx}
\usepackage{multirow}

\newcommand{\PreserveBackslash}[1]{\let\temp=\\#1\let\\=\temp}

\begin{document}

\title{Green's Function Reaction Dynamics: a new approach to simulate
biochemical networks at the particle level and in time and space}

\author{Jeroen S. van Zon}
\affiliation{Division of Physics and Astronomy, Vrije Universiteit, De
Boelelaan 1081, 1081 HV Amsterdam, The Netherlands.}
 
\author{Pieter Rein ten Wolde}
\affiliation{FOM Institute for Atomic and Molecular Physics,
Kruislaan 407, 1098 SJ Amsterdam, The Netherlands.}
\affiliation{Division of Physics and Astronomy, Vrije Universiteit, De
Boelelaan 1081, 1081 HV Amsterdam, The Netherlands.}

\date{\today}

\begin{abstract}
Biochemical networks are the analog computers of life. They allow living cells to control a large number of biological processes, such as
gene expression and cell signalling. In biochemical networks, the
concentrations of the components are often low. This means that the
discrete nature of the reactants and the stochastic character of their
interactions have to be taken into account. Moreover, the spatial
distribution of the components can be of crucial importance. However,
the current numerical techniques for simulating biochemical networks
either ignore the particulate nature of matter or treat the spatial
fluctuations in a mean-field manner.  We have developed a new
technique, called Green's Function Reaction Dynamics (GFRD), that
makes it possible to simulate biochemical networks at the particle
level and in both time and space. In this scheme, a maximum time step is
chosen such that only single particles or pairs of particles have to
be considered. For these particles, the Smoluchowski equation can be
solved analytically using Green's functions. The main idea of GFRD is
to exploit the exact solution of the Smoluchoswki equation to set up
an event-driven algorithm. This allows GFRD to make large jumps in
time when the particles are far apart from each other. Here, we apply
the technique to a simple model of gene expression. The simulations reveal that
the scheme is highly efficient. Under biologically relevant
conditions, GFRD is up to six orders of magnitude faster than
conventional particle-based techniques for simulating biochemical
networks in time and space.
\end{abstract}


\maketitle 

\section{Introduction}

Organisms can be viewed as information
processing machines. Even relatively simple organisms, such as the
bacterium {\em Escherichia coli}, can perform fairly complex
computations such as: \texttt{if lactose is present and not glucose is present,
then use lactose}. Recent technological developments
have made it possible to obtain information on the regulatory
architecture of the cell on an unprecedented scale. In addition, extensive
databases are now available that catalog biochemical pathways. Nevertheless,
our understanding of the mechanisms that allow living cells to process
information  is still limited. One important reason for this is that
these mechanisms are controlled by stochastic processes.

In the living cell, computations are performed by molecules that
chemically and physically interact with each other. These components,
that form what is called a biochemical network, behave
stochastically. They often move in an erratic fashion, namely by
diffusion, and act upon each other in a stochastic manner - chemical
reactions, and equally important, physical interactions are
probabilistic in nature. These factors become particularly important
when the concentrations are low. In the living cell, this is often the
case, and, as a result, biochemical networks can be highly
stochastic~\cite{McAdams97,Elowitz00}. In this respect, it is a
remarkable fact that many biological processes operate reliably with
surprisingly small numbers of molecules.

Another important reason for our limited understanding of biochemical
networks is that they operate not only in time, but also in
space. In the living cell, signals often have to be transmitted from
one place to the next by the diffusive motion of ``messenger''
molecules; their concentrations can be non-uniform, and more
importantly, their low mobility can limit the response time of the
network. Moreover, many processes, such as the immune response,
involve a complex spatial reorganization of the reactants. Finally, a
large number of biological activities require the localized assembly of
a complex of proteins. All these processes can only be accurately
investigated using techniques that resolve the network in time and space.

In principle, computer simulations are ideally suited for studying how
biochemical networks reliably process information in time and
space. However, the current numerical techniques are of limited use
for this purpose. Table \ref{tab:overview} gives an overview of the
commonly used techniques for analysing biochemical networks.  The
conventional approach is to write down the macroscopic rate equations
and to solve the corresponding differential equations numerically. In
this method, the evolution of the network is deterministic. It is
implicitly assumed that the concentrations are large and that
fluctuations can be neglected. The effect of fluctuations is often
included by adding a noise term to the macroscopic rate equations
\cite{VanKampen}. However, at low concentrations, this approach is
bound to fail, as demonstrated by Togashi and Kaneko~\cite{Togashi}
and Shnerb and coworkers~\cite{Shnerb}.  At low concentrations, we
have to recognize the discrete nature of the reactants and the
stochastic character of their interactions. Currently, two techniques
exist for simulating biochemical networks at the particle
level~\cite{Gillespie76,Bray}. Both of these are consistent with the
chemical master equation. However, the chemical master equation relies
on the assumption that there are many non-reactive collisions to stir
the system between the reactive collisions. In effect, it is
implicitly assumed that at each instant the particles are uniformly
distributed in space. This is a serious limitation. As discussed
above, the functioning of many biochemical networks depends crucially
on their spatial organization. Moreover, fluctuations of the
components in space can be a major source of noise in biochemical
networks.

\let\PBS=\PreserveBackslash
\begin{table*}[t]
\begin{tabular}{>{\PBS\raggedright\hspace{0pt}}p{30mm}|>{\PBS\raggedright\hspace{0pt}}p{30mm}|>{\PBS\centering\hspace{0pt}}p{30mm}|>{\PBS\centering\hspace{0pt}}p{30mm}}
Description & & Accounts for spatial extent of network & Incorporates fluctuations \\ \hline \vspace*{0.2cm}
\multirow{3}{40mm}[0mm]{Continuum} & Ordinary differential equations &
No & No\\ \cline{2-4}
& Stochastic differential equations & No & Only at high
concentrations\\\cline{2-4}
& Reaction diffusion equations & Yes & No \\\hline
\multirow{2}{40mm}[0mm]{Particle-based} & Chemical master equation & No & Yes \\\cline{2-4}
& GFRD & Yes & Yes
\end{tabular}
\caption{\label{tab:overview} Overview of the commonly used techniques and
the newly developed technique, called Green's Function Reaction
Dynamics (GFRD), to simulate biochemical networks. GFRD takes both the
discrete nature and the spatial distribution of the reactants into
account.}
\end{table*}

Clearly, in many cases we have to describe biochemical networks both
temporally and spatially at the molecular level. However, the current
techniques for simulating biochemical networks cannot accomplish this
since they either ignore the particulate nature of matter, or treat
the spatial fluctuations in a mean-field manner. We have developed a
new technique, named Green's Function Reaction Dynamics (GFRD), that
makes it possible to simulate biochemical networks at the particle
level and in time and space. This technique, which we describe in
detail in the next section, is an event-driven algorithm in which the
particles are propagated in space between reaction events using
Green's functions.  In the subsequent section, we apply GFRD to a
simple model of gene expression. The calculations show that the
technique is highly efficient. We thus believe that GFRD brings
simulating biochemical networks at the particle level and in time and
space within reach.

\section{Numerical technique}

\subsection{Introduction}
\label{sec:tech_intro}
In this section we discuss the main ideas underlying Green's Function
Reaction Dynamics (GFRD). GFRD is a stochastic  scheme that
combines the propagation of the particles in space with the
reactions between them in an efficient manner. Most proteins are
believed to be transported by their diffusive motion, although in
eukaryotic cells, cytoskeletal networks and motor proteins may
facilitate the transport of molecules~\cite{Albertsbook}. Such
mechanisms of active transport have not been observed in bacteria,
where diffusion is believed to be the primary means of intracellular
movement. For now, we will assume that the network components are
transported by diffusion, although the technique could be extended to
include active transport as well.

Two approaches seem to be potentially useful for simulating  biochemical
networks at the particle level and in time and space. The first is to
let the particles undergo a random walk on a lattice and to let
reaction partners react with a certain probability when they happen to
meet each other. This technique has a number of limitations, the most
important of which are that the physical dimensions of the particles
and the interactions between them cannot conveniently be described.

Brownian Dynamics is a more appealing technique. This is
a stochastic dynamics scheme, in which the particles are propagated
in space according to the overdamped limit of the Langevin equation. In
Brownian Dynamics, the solvent is considered implicitly; only the
solute particles are considered explicitly.  The forces experienced by
these particles contain two parts: a conservative part, which arises
from the interactions with the other solute particles, and a random
part.  The latter is the dynamical remnant of the solvent - the
solutes are thought to experience random forces from the solvent. Via
the fluctuation-dissipation theorem and the Einstein relation, the
random forces are related to the diffusion constant of the
particles. To be more explicit, the equations of motion for the solute
particles are given by:
\begin{equation} 
\label {eq:BD} 
\dot{{\bf r}}_s = \frac{D_s}{k_B T}
\left( {\bf F}_s + \delta {\bf F}_s \right).  
\end{equation} 
Here, ${\bf r}_s$ denotes the position of solute particle $s$, $D_s$
is the diffusion constant of solute particle $s$, $k_B T$ is
Boltzmann's constant times temperature, ${\bf F}_s$ is the
conservative force exerted by the other solute particles, and $\delta
{\bf F}_s$ is the random force that arises from the interactions with
the solvent. 

Brownian Dynamics has a number of advantages over lattice-based
techniques: the particles move in continuum space; the interactions
between particles - the potential of mean force - can easily be
described; excluded volume effects are taken into account naturally;
and  a different diffusion constant can be
assigned to each type of particles. 

In principle, chemical reactions can be
straightforwardly implemented into the Brownian Dynamics scheme: the
particles are propagated according to Eq.~(\ref{eq:BD}) and when two
reaction partners happen to meet each other, they can react with a
probability that is consistent with the rate constant.  However,
the major drawback of such a scheme is that very small time
steps are needed in order to resolve the collision events. This makes
brute-force Brownian Dynamics a very inefficient scheme to simulate
biochemical networks.

The main idea of GFRD is to combine in one step the propagation of the
particles in space with the reactions between them. To see how this
can be accomplished, one should realize that Brownian Dynamics is, in
effect, a {\em numerical} procedure for solving the Smoluchowski
equation~\cite{Smoluchowski17}. However, for a pair of particles, that not only
move diffusively, but also can react according to $A+B \longrightarrow
C + D + ...$, the Smoluchowski equation can be solved {\em
analytically} using Green's functions. The Green's function for the
pair of particles $A$ and $B$, $p({\bf r},t|{\bf r_0},t_0)$, yields
the probability that the inter-particle vector ${\bf r}_0$ at time
$t_0$ becomes ${\bf r}$ at a later time $t$. The essence of GFRD is to
exploit this exact solution for a pair of particles to set up an
event-driven algorithm. This allows GFRD to make large jumps in time
when the particles are far apart from each other. In biochemical
networks this is often the case as the reactant concentrations are
usually low. GFRD is therefore ideally suited for biochemical
networks. Below we describe the scheme in more detail.

\subsection{Overview of the algorithm}
\label{sec:overview}

\begin{figure}[b] \centering
\includegraphics[width=8cm]{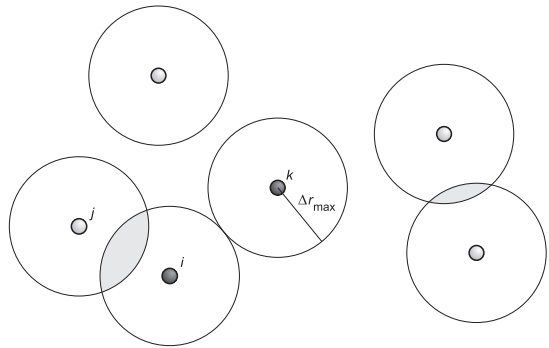} \caption{Determination of the maximum
time step, $\Delta t_{\rm max}$. The maximum size of the time step is
set by the requirement that each particle can interact with at most
one other particle during that time step; each
particle $i$ can thus travel a distance of at most $\Delta r_{{\rm max},i}$
during a time step, as indicated by the circles. We
have used the operational criterion $\Delta r_{{\rm max},i} = H
\sqrt{6 D_i \Delta t_{{\rm max},i}}$, with $D_i$ being the
diffusion constant of particle $i$. A value of $H=3$ was found to yield a
good conservation of the correct steady-state distribution. In this
example, each particle is assumed to have the same diffusion
constant; the time step is limited by the constraint that particles
$i$ and $k$ should not interact as particle $i$ can already interact
with particle $j$. Note that with this maximum time step the
many-body problem of propagating the $N$ particles is reduced to that
of propagating single particles and pairs of particles.\label{fig:overview}}
\end{figure}

Let us imagine a configuration of reactants as shown in
Fig.~\ref{fig:overview}. The circles indicate the distances the
particles can travel in a time step $\Delta t$. For a particle that
moves by free diffusion with a diffusion constant $D$, that distance
scales as $\langle |\Delta {\bf r}|^2\rangle \sim D \Delta
t$. Intersecting circles represent particles that may meet within the
time step $\Delta t$. Isolated particles and pairs of interacting
particles can be propagated analytically using Green's functions, as
we will discuss in detail in the next section. Clearly, the larger the time
step, the larger the circles and the greater the probability that
reaction partners meet and react with each other.  However, we cannot
make the time step arbitrarily large: if a given particle can collide
with more than one other particle during a time step, then propagating
the particles becomes a many-body problem that we can not solve
analytically. The size of a time step is thus limited by the
requirement that each particle can interact with at most one other
particle during that time step. This constraint sets an upper limit on
the magnitude of a time step in our algorithm; we will call it $\Delta
t_{\rm max}$. However, provided that we consider times $\Delta t$
smaller than $\Delta t_{\rm max}$, the problem can be reduced to that
of propagating single particles and pairs of particles. This can be
solved analytically using Green's functions, as we will now describe.

\subsection{Monomolecular reactions - the Green's function for single particles}
In this section, we consider the propagation of a {\em single}
particle. We assume that the particle is spherical in shape and moves
by free diffusion with a diffusion constant $D$. The diffusive motion
of the particle is described by the Einstein diffusion equation
\begin{equation}
\partial_t p_1({\bf r},t|{\bf r}_0,t_0) = D \nabla^2 p_1({\bf r},t|{\bf
r}_0,t_0).
\end{equation}
Here, $p_1({\bf r},t|{\bf r}_0,t_0)$ is the probability that the
particle is at position ${\bf r}$ at time $t$, given that is was at
${\bf r}_0$ at time $t_0$. This diffusion equation can be solved
subject to the initial condition and the boundary condition
\begin{eqnarray}
p_1({\bf r},t_0|{\bf r}_0,t_0) &=& \delta ({\bf r}-{\bf r}_0),\\
p_1({\bf |r|}\rightarrow \infty,t|{\bf r}_0,t_0) &=& 0,
\end{eqnarray}
respectively. The solution $p_1({\bf r},t|{\bf r}_0,t_0)$ is known as
a Green's function. It is given by the well-known expression
\begin{equation}
\label{eq:free_diff}
p_1({\bf r},t|{\bf r}_0,t_0) =
\frac{1}{\left[4 \pi D (t-t_0)\right]^{3/2}} \exp \left[-\frac{|{\bf r}-{\bf
r}_0|^2}{4 D (t-t_0)}\right].
\end{equation}

We now consider the case in which the particle  does not only move
diffusively, but also can ``decay'' according to
\begin{equation}
\label{eq:reactdecay}
A\stackrel{k_d} \longrightarrow B + C \dots
\end{equation}
We will assume that, if the reaction happens, it happens {\em
instantaneously}. This means that the reaction can be decoupled from
the diffusive motion of the particle. If the reaction is a Poisson
process with $k_d dt$ being the probability
that a reaction occurs in an infinitesimal time interval $dt$, then
the probability that the {\em next} reaction occurs between $t$ and
$t+dt$, is
\begin{equation}
\label{eq:nrd}
q_d(t|t_0) dt = k_d \exp \left[-k_d (t-t_0)\right] dt.
\end{equation}
In section~\ref{sec:algorithm}, we will use Eqs.~\ref{eq:free_diff}
and~\ref{eq:nrd} to set up an event-driven algorithm.

\subsection{Bimolecular reactions - the Green's function for pairs of particles}
In this section, we consider {\em one pair} of particles $A$ and $B$
that can react according to
\begin{equation}
A+B \stackrel{k_a} \longrightarrow C + D \dots
\end{equation}
We again assume that the particles $A$ and $B$ are spherical in shape and
move by their diffusive motion; the diffusion constants for particle
$A$ and $B$ are $D_A$ and $D_B$, respectively. Furthermore, we assume
that the particles react with an intrinsic rate constant $k_a$ when they
have approached each other within the reaction distance
$\sigma=(d_A+d_B)/2$, where $d_A$ and $d_B$ are the diameters of
particles $A$ and $B$, respectively. In addition, we imagine that the
particles interact with each other via a potential $U({\bf r})$, where
${\bf r} = {\bf r}_B - {\bf r}_A$. The force acting on particle B is
thus given by $-\nabla_B U({\bf r}) = {\bf F(r)}$, while the force
acting on particle A is given by $-{\bf F(r)}$.

We aim to derive the distribution function $p_2({\bf r}_A,{\bf
r}_B,t|{\bf r}_{A0},{\bf r}_{B0},t_0)$, which yields the probability
that the particles $A$ and $B$ are at positions ${\bf r}_{A}$ and
${\bf r}_{B}$ at time $t$, given that they were at ${\bf
r}_{A0}$ and ${\bf r}_{B0}$ at time $t_0$. This distribution function
satisfies for ${\bf |r|}\geq\sigma$ the following Smoluchowski equation~\cite{Smoluchowski17}
\begin{eqnarray}
\label{eq:Smol}
\lefteqn{
\partial_t p_2({\bf r}_A,{\bf
r}_B,t|{\bf r}_{A0},{\bf r}_{B0},t_0) =} \nonumber \\
&& [D_A \nabla^2_A + D_B \nabla^2_B 
	 - D_B \beta \nabla_B \cdot {\bf F(r)} + D_A \beta 
	\nabla_A \cdot {\bf F(r)}] \nonumber \\
&& {} \times p_2({\bf r}_A,{\bf r}_B,t|{\bf r}_{A0},{\bf r}_{B0},t_0).
\end{eqnarray}
It will be convenient to make a coordinate transformation
\begin{eqnarray}
{\bf R} &=& \sqrt{D_B/D_A} {\bf r}_A + \sqrt{D_A/D_B} {\bf r}_B,\label{eq:RAB}\\
{\bf r} &=& {\bf r}_B - {\bf r}_A \label{eq:rAB},
\end{eqnarray}
and to define the operators
\begin{eqnarray}
\nabla_{\bf R} &=& \partial/\partial {\bf R},\\
\nabla_{\bf r} &=& \partial/\partial {\bf r}.
\end{eqnarray}
Eq.~\ref{eq:Smol} can then be rewritten as:
\begin{eqnarray}
\label{eq:SmolRr}
 \partial_t p_2({\bf R},{\bf r},t|{\bf R}_{0},{\bf r}_{0},t_0) 
	&=& (D_A \! + \! D_B)[ \nabla_{\bf R}^2 \! + \! \nabla_{\bf r}\! \cdot \!(\nabla_{\bf r} \!-\! {\bf F(r)})] \nonumber \\
&& {} \times p_2({\bf R},{\bf r},t|{\bf R}_{0},{\bf r}_{0},t_0), \nonumber\\
&& {\bf |r|}\geq \sigma.
\end{eqnarray}
It is seen that Eq.~\ref{eq:SmolRr}
describes two independent random processes - free diffusion in the
coordinate ${\bf R}$ and diffusion with a drift in the coordinate
${\bf r}$. This means that the distribution function $p_2({\bf r}_A,{\bf
r}_B,t|{\bf r}_{A0},{\bf r}_{B0},t_0)$ can be factorized as
$p_{2}^{\bf R}({\bf
R},t|{\bf R}_0,t_0)p_{2}^{\bf r}({\bf r},t|{\bf r}_0,t_0)$ and that the above
equation can be reduced to one Smoluchowski equation for the
coordinate ${\bf R}$ and one for the coordinate ${\bf r}$:
\begin{eqnarray}
\partial_t p_{2}^{\bf R}({\bf R},t|{\bf R}_0,t_0) &=&(D_A + D_B) \nabla_{\bf R}^2 \nonumber \\
&&{} \times p_{2}^{\bf R}({\bf R},t|{\bf R}_0,t_0), \label{eq:pR}\\ 
\partial_t p_{2}^{\bf r}({\bf r},t|{\bf
	r}_0,t_0) &=& (D_A + D_B) \nabla_{\bf r} \cdot (\nabla_{\bf r}
	- {\bf F(r)})  \nonumber\\
&& {} \times p_{2}^{\bf r}({\bf r},t|{\bf r},t_0), {\bf |r|}\geq \sigma . \label{eq:pr}
\end{eqnarray}
Eqn.~\ref{eq:pR} describes the free diffusive motion of the coordinate
${\bf R}$. The solution of that equation, for the initial condition
$p_2^{\bf R}({\bf R},t_0|{\bf R}_0,t_0) = \delta ({\bf
R}-{\bf R}_0)$ and boundary condition $ p_2^{\bf R}({\bf |R|}\rightarrow
\infty,t|{\bf R}_0,t_0) = 0$, is
\begin{equation}
\label{eq:pGR}
p_2^{\bf R}({\bf R},t|{\bf R}_0,t_0) = \frac{\exp \left[-\frac{|{\bf R}-{\bf R}_0|^2}{4
(D_A+D_B) (t-t_0)}\right]}{\left[4 \pi (D_A+D_B)
(t-t_0)\right]^{3/2}} .
\end{equation}

The non-trivial solution is that of the Smoluchowski equation for the
inter-particle vector ${\bf r}$. This solution also has to take into account the
reactions between $A$ and $B$. We will incorporate the reaction as a
boundary condition on the solution of the Smoluchoswki equation. To be
more explicit, the initial condition and boundary conditions for
the coordinate ${\bf r}$ are given by
\begin{eqnarray}
p_2^{\bf r}({\bf r},t_0|{\bf r}_0,t_0) &=& \delta ({\bf r}-{\bf
r}_0),\\ 
\label{eq:bca_infty}
p_2^{\bf r}({\bf |r|}\rightarrow \infty,t|{\bf r}_0,t_0) & = & 0,\\
-j(\sigma,t|{\bf r}_0,t_0) &\equiv&4\pi \sigma^2 D \left(\frac{\partial}{\partial r} - {\bf F}({\bf r})\right) \nonumber \\
&& {} \times p_2^{\bf r}({\bf r},t|{\bf r}_0,t_0)|_{{\bf |r|}=\sigma}, \nonumber\\
 &=&k_a p_2^{\bf r}({\bf |r|}=\sigma,t|{\bf r}_0,t_0)\label{eq:radbc},
\end{eqnarray}
where $\partial /\partial_r$ denotes a derivative with respect to the
inter-particle separation $r$.  It is seen that the reaction enters the
problem as a third boundary condition on the solution of the
Smoluchowski equation. Here $j(\sigma,t|{\bf r}_0,t_0)$ is the outward
radial flux of probability $p_2^{\bf r}({\bf r},t|{\bf r}_0,t_0)$ through the
``contact'' surface of area $4\pi \sigma^2$. The boundary condition,
also known as a {\em radiation} boundary
condition~\cite{Carslaw,Agmon90}, states that this radial flux of
probability equals the intrinsic rate constant times the probability
that the particles $A$ and $B$ are, in fact, in contact. In the limit
$k_a \rightarrow \infty$, the radiation boundary condition reduces to
an {\em absorbing} boundary condition $p_2^{\bf r}({\bf |r|}=\sigma,t|{\bf
r}_0,t_0)=0$, while in the limit $k_a \rightarrow 0$ the radiation
boundary condition reduces to a {\em reflecting} boundary condition.

The Green's function $p_2^{\bf r}({\bf r},t|{\bf r}_0,t_0)$ is derived in
appendix A for the case in which ${\bf F(r) = 0}$ for $|{\bf
r}|>\sigma$; for cases in which ${\bf F(r) \neq 0}$ for $|{\bf
r}|>\sigma$, the Green's functions could, depending upon the
interaction potential, either be obtained analytically or
numerically. Here, we will discuss some useful quantities that can be
derived from the Green's function. The first quantity of interest is
the probability that a pair of particles, initially separated by ${\bf
r}_0$, survives and does not recombine by time $t$. This so-called
{\em survival} probability is given by
\begin{equation}
\label{eq:Sa}
S_a(t|{\bf r}_0,t_0) = \int_{\bf |r|>\sigma} d{\bf r} p_2^{\bf r}({\bf r},t|{\bf
r}_0,t_0).
\end{equation}
Clearly, $S_a(0|{\bf r}_0,t_0)=1$ for $|{\bf r}_0|>\sigma$. The second
quantity of interest is the reaction rate $q_a(t|{\bf r}_0,t_0)$, which
is defined as the probability per unit time that a pair, initially
separated by ${\bf r}_0$, reacts at time $t$. It is
related to the survival probability by
\begin{equation}
\label{eq:nra}
q_a(t|{\bf r}_0,t_0) \equiv -\frac{\partial S_a(t|{\bf r}_0,t_0)}{\partial t}.
\end{equation} 
Since the reactions are assumed to occur only at contact, the reaction
rate is also given by the flux at contact
\begin{equation}
q_a (t|{\bf r}_0,t_0) = -j(\sigma,t|{\bf r}_0,t_0).
\end{equation}
The above equation can also be obtained from Eq.~\ref{eq:pr} and
Eq.~\ref{eq:Sa} and by using the fact that the flux at ${\bf
|r|}\rightarrow\infty$ vanishes.

The reaction rate $q_a(t|{\bf r}_0,t_0)$ can be interpreted as the
probability that the {\em next} reaction for a pair of
particles, initially separated by ${\bf r}_0$, occurs at time
$t$. This is used to set up the GFRD event-driven algorithm, which we
will describe in the next section.

\subsection{Outline of the algorithm}
\label{sec:algorithm}
To explain the essence of the algorithm, it will be instructive to
consider a single particle of type $A$ that can react with a single
particle of type $B$
according to the following scheme
\begin{equation}
A + B \rightleftarrows C.
\end{equation}
Furthermore, it will be useful to imagine that these particles are
surrounded by neighbors, the presence of which limit the size of the
time step to $\Delta t_{\rm max}$. As a function of time, the system
will flip-flop between the associated state $C$ and the dissociated
state $A+B$. The GFRD event-driven algorithm to propagate this system would
consist of iterating the following steps: 

\begin{enumerate}
{\item
 If the system is in the
dissociated state $A+B$, then draw a next association time $t$
according to $q_a(t|{\bf r}_0,t_0)$ (Eq.~\ref{eq:nra}). }

{\subitem
 a) If $(t-t_0)
\geq \Delta t_{\rm max}$, then the two particles will not react within
the time step; new positions for $A$ and $B$ at time $t_0+\Delta
t_{\rm max}$ are obtained from $p_2^{\bf R}({\bf R},t_0+\Delta t_{\rm
max}|{\bf R}_0,t_0)$ (Eq.~\ref{eq:pGR}) and $p_2^{\bf r}({\bf
r},t_0+\Delta t_{\rm max}|{\bf r}_0,t_0)$ (Eq.~\ref{eq:pGr}) with
${\bf R}$ and ${\bf r}$ as given by Eqs.~\ref{eq:RAB}
and~\ref{eq:rAB}. }
{\subitem 
b) If $(t-t_0) < \Delta t_{\rm max}$, then the next
reaction will occur within the time step; a new position for particle
$C$ at time $t$ is obtained from $p_2^{\bf R}({\bf R},t|{\bf
R}_0,t_0)$ (Eq.~\ref{eq:pGR}). 
}
{\item 
 If the system is in the associated
state $C$, then draw a next dissociation time from $q_d(t|t_0)$
(Eq.~\ref{eq:nrd}). }

{\subitem 
a) If $(t-t_0) \geq \Delta t_{\rm max}$, then
particle $C$ will not have decayed by $t_0+\Delta t_{\rm max}$; a new
position for particle $C$, ${\bf r}_C$, at time $t_0+\Delta t_{\rm
max}$ is obtained from $p_1({\bf r}_C,t_0+\Delta t_{\rm max}|{\bf
r}_{C0},t_0)$ (Eq.~\ref{eq:free_diff}); }
{\subitem
 b) If $(t-t_0) < \Delta t_{\rm
max}$, the next reaction will occur within the maximum time step; the
particles $A$ and $B$ are placed at time $t$ adjacent to each other at
positions around ${\bf r}_C$ as obtained from $p_1({\bf r}_C,t|{\bf
r}_{C0},t_0)$ (see Eq.~\ref{eq:free_diff}).}

\end{enumerate}

The procedure outlined above forms the heart of the algorithm. The
crux of the method is to choose the maximum time step such that only
monomolecular reactions or bimolecular reactions have to be considered. This
makes it possible to use the exact solution of the Smoluchowski
equation to propagate the system to the next reaction event in a
single step.  The full algorithm for a system of $N$ particles thus
consists of iterating the following steps:

\begin{enumerate}

\item
\underline{Determine maximum time step $\Delta t_{\rm max}$.} The
maximum time step is determined by the condition that only single particles
or pairs of particles have to be considered (see
section~\ref{sec:overview} and Fig.~\ref{fig:overview}).  
For each particle $i$, we determine the maximum time step $\Delta
t_{{\rm max},i}$, such that it can interact with at
most one other particle. The maximum global time step is then given by
\begin{equation}
\Delta t_{\rm max} = \mbox{min}(\{\Delta t_{{\rm max},i}\}).
\end{equation}
In order to determine $\Delta t_{{\rm max},i}$ for particle $i$, we
assume that during that step the particle can
travel at most a distance $\Delta r_{{\rm max},i} = H \sqrt{6 D_i
\Delta t}$, where $D_i$ is the diffusion constant of particle $i$. We
find that  $H=3$  suffices to preserve the correct steady-state distribution.

\item
\underline{Determine next reaction and next reaction time}.  We first
construct a list of possible reactions $\{R_\nu\}$.  With
each reaction $R_\nu$, we associate a survival probability
function $S_\nu(t-t_0)$ and a next-reaction distribution function $q_\nu
(t-t_0)$; the two are related via $q_\nu(t-t_0) = -\partial S_\nu(t-t_0)/\partial
t$.  For the bimolecular reactions, $q_\nu (t-t_0) = q_a(t|{\bf r}_0,t_0)$ as
given by Eq.~\ref{eq:nra} and $S_\nu(t-t_0) = S_a(t|{\bf r}_0,t_0)$ as
given by Eq.~\ref{eq:Sa}. For the monomolecular reactions, $q_\nu (t-t_0) =
q_d(t|t_0) = k_d \exp(-k_d (t-t_0))$ and $S_\nu(t-t_0) = \exp(-k_d
(t-t_0))$.

For each reaction $R_\nu$, we generate a random number $\xi_\nu$,
 uniformly distributed in the interval $0<\xi_\nu<1$.  If $\xi_\nu
 \leq (1-S_\nu(\infty))$, a tentative next reaction time $\Delta t_\nu$ is
 obtained from
\begin{equation}
\xi_\nu = \int_0^{\Delta t_\nu} q_\nu(t') dt' = 1 - S_\nu (\Delta t_\nu).
\end{equation}
If, however, $\xi_\nu > (1- S_\nu(\infty))$, then the reaction $R_\nu$
does not occur and it is dropped from the list of possible reactions.
From the remaining list of tentative reactions, we choose as the
actual next reaction the one that occurs first, {\em provided} that
this reaction occurs within the maximum time step $\Delta t_{\rm
max}$. Accordingly, the system will be propagated through a time $\Delta
t$ as given by
\begin{equation}
\Delta t = {\rm min} (\{\Delta t_\nu\},\Delta t_{\rm max}).
\label{eq:t_prop}
\end{equation}
Note that if there is no tentative reaction for which the tentative
next reaction time $\Delta t_\nu < \Delta t_{\rm max}$, then no reaction will
occur within the time step.  Here, we also mention for completeness that
for association reactions $S_\nu(\infty) \neq 0$: for two particles,
that can diffuse and react subject to the boundary condition
Eq.~\ref{eq:bca_infty}, there is a finite probability that they {\em
never} react; this is related to the well-known fact that a random
walker, that starts at the origin, can ``escape'' and never return to
the origin.

\item
\underline{Propagate particles.} The single particles are propagated
according to $p_1({\bf r},t|{\bf r}_0,t_0)$ in Eq.~\ref{eq:free_diff};
if a particle decays, then the products are placed next to each other
at positions centered around ${\bf r}$. For each pair of particles, the
following two substeps are executed: 1) a new position for the
coordinate ${\bf R}$ is obtained from Eq.~\ref{eq:pGR}; 2) if the pair
has not reacted, a new inter-particle vector ${\bf r}$ is obtained
from $p_2^{\bf r}({\bf r},t|{\bf r}_0,t_0)$ in Eq.~\ref{eq:pGr}; else,
if it has reacted, the products are placed adjacent to each other at
positions around ${\bf R}$.

\item
\underline{Update particles.} Update identities of particles according
to the executed reaction. Delete or add particles created or destroyed
in that reaction. 
 
\end{enumerate}
A proof of the algorithm is given in appendix B.

Finally, we remark that the assumptions made above, namely that
particles are spherical in shape and move by (free) diffusion, are not
essential. An event-driven algorithm of this type could be set up in
the case of non-spherical particles, interacting particles and/or
those moving by other mechanisms than diffusion such as active
transport. In these cases, the required Green's functions could be
obtained numerically if necessary.

\section{Results}
This section is organized as follows: first we study a simple
bimolecular reaction to show that GFRD accurately reproduces
theoretical results. Then we turn our attention to a very simple
model of gene expression as a typical example of a system that is well
handled by the GFRD technique. We specifically focus on the levels of
noise in protein concentrations and find dramatic differences between
GFRD and results from the chemical master equation, that ignores
spatial fluctuations. Finally we compare the performance of GFRD to a
conventional Brownian Dynamics algorithm.

\subsection{Bimolecular reaction}
\label{sec:bimol}
To test the validity of our approach, we study the reversible bimolecular reaction
\begin{equation}
A + B \overset{k_a}{\underset{k_d}\rightleftarrows} C
\label{eq:bimolreact}
\end{equation}
with forward rate constant $k_a$ and backward rate constant $k_d$. As
a first test of our algorithm, we focus on an {\em isolated} pair of
particles $A$ and $B$. We use a set up in which particle $A$,
with diameter $\sigma$, is placed at the origin and held fixed during
the simulation. The second particle $B$, also with diameter $\sigma$, is
initially placed at random in a spherical shell of radius $r_0$
centered around particle $A$. We then propagate particle $B$ for a
time $t_{\rm sim}$. During this time, particle $B$ can diffuse freely
with a diffusion constant $D$ and it can associate with particle $A$
with a rate constant $k_a$ and dissociate from it with a rate constant
$k_d$. Typically, particle $B$ will associate with and dissociate from
$A$ a (large) number of times during the simulation. We repeat this whole
procedure many times. This allows us to calculate the distribution
function $p_{\rm rev} (r,t|r_0,t_0)$, which gives the probability that
the two particles $A$ and $B$, separated by a distance $r_0$ at time
$t_0$, are a distance $r$ apart at a later time $t$. This numerical
result can be compared to the analytical solution recently derived by
Kim and Shin for the {\em reversible} reaction shown in Eq.~\ref{eq:bimolreact}~\cite{Kim}.

\begin{figure}[b] \centering
\includegraphics[width=7.5cm]{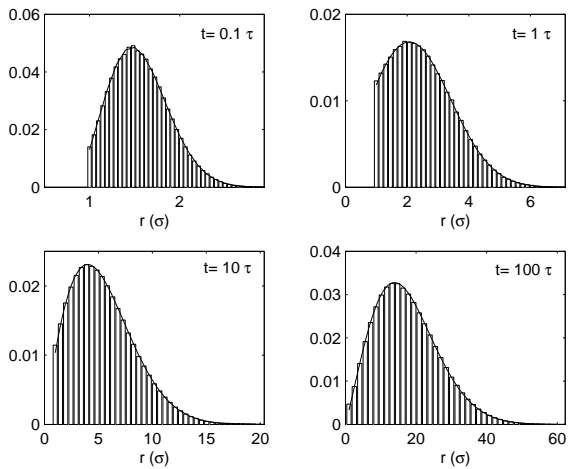} \caption{\label{fig:Gf} The distribution
$p_{\rm rev}(r,t|\sigma,t_0=0)$ for $t=0.1\tau$, $1\tau$, $10\tau$ and $100\tau$. For
$r<\sigma$, the distribution $p_{\rm rev}(r,t|\sigma,t_0)=0$ due to the hard
sphere repulsion between the particles.  The bars denote the
simulation results and the solid lines denote the analytical solutions of
Kim and Shin~\cite{Kim}. Note that the particles are
initially placed at contact ($r_0 = \sigma$).\label{fig:GFcompare} The
forward rate constant $k_a = 1000 \mbox{ molecule}^{-1} \sigma^3
\tau^{-1}$ and the backward rate constant is $k_d = 1 \tau^{-1}$. The
unit of time $\tau=\sigma^2/D$.}
\end{figure}

If the next reaction time were larger than the simulation time,
$t_{\rm sim}$, then we could in principle directly propagate the
particles through $t_{\rm sim}$. However, this would not provide a
stringent test of our algorithm. At each step, we therefore choose a
maximum time step $\Delta t_{\rm max}$ at random from the interval
$[10^{-4} \tau,t_{\rm sim}]$, where $\tau = \sigma^2/D$ is the unit of
time. This could be interpreted as mimicking the constraint on the
maximum time step arising from the presence of other particles.

Figure \ref{fig:GFcompare} shows excellent agreement for $p_{\rm
rev}(r,t|r_0,t_0)$ between theory and simulation for $r_0=\sigma$. We
find similar agreement between theory and simulation for other initial
distances $r_0$ and for other values of the diffusion constant $D$ and
reaction rates $k_a$ and $k_d$. It should be realized that because the
particles are initially placed at contact, many reactions can occur
during the time $t_{\rm sim}$. Moreover, because we divide the
simulation time into smaller intervals, we must propagate the
particles many times, using the Green's function for an extensive
range of ${\bf r}$, $t-t_0$ and ${\bf r}_0$. Thus, at least for the
case of an isolated pair of particles, this procedure provides a
thorough test of our algorithm.

Next, we want to study a more complex system in which a single
particle of type $A$ is held fixed at the center of a spherical container of
radius $R$ and is surrounded by $N_B$ particles of type $B$.  Particle $A$ can
again react with a particle $B$ to form the product $C$ according to
the scheme in Eq.~\ref{eq:bimolreact}; particle $C$, like particle $A$,
does not diffuse. Particles $B$ and $C$ do not react, although
they are not allowed to overlap with each other. The excluded volume
interactions between a pair of two $B$ particles and between a pair of a $B$
and a $C$ particle is taken into account by using reflecting boundary
conditions, i.e. by setting $k_a = 0$ in Eq.~\ref{eq:radbc}. We note
that the requirement that the $B$ and $C$ particles are not allowed to
overlap, may impose a constraint on the maximum size of the time step,
$\Delta t_{\rm max}$. The wall of the container is assumed to be
reflecting.  As no analytical solution exists for a {\em pair} of
particles in the presence of a reflecting boundary, we introduce the
further requirement that during a time step a particle can only
interact with either the wall of the container or with another
particle, but not with both.

As the $B$ particles diffuse through the container, they will come
into contact with the fixed particle $A$. When in contact, the
particles $A$ and $B$ can enter the \emph{bound} state $C$ with
forward rate $k_a$. When in the bound state, other $B$ particles
approaching the fixed $C$ particle cannot react with it. Only after
dissociation into the unbound state $A+B$, occurring at rate
$k_d$, can another reaction occur. On average, there will be a
probability $p_{\text{bound}}$ of finding the $A$ particle
bound to the $B$ particle. It is given by
\begin{equation}
\label{eq:pbound}
p_{\text{bound}} = \frac{K N_B}{V + K N_B},
\end{equation} 
where
$V$ is the volume available for the $B$
particles, $N_B$ is the \emph{total} number of $B$ particles and
\begin{equation}
\label{eq:Keq}
K=g_{AB}(\sigma)k_a/k_d
\end{equation} 
is the equilibrium constant. The
function $g_{AB} (r)$ is the radial distribution function for the pair
of particles $A$ and $B$.

\begin{figure}[b] \centering
\includegraphics[width=8cm]{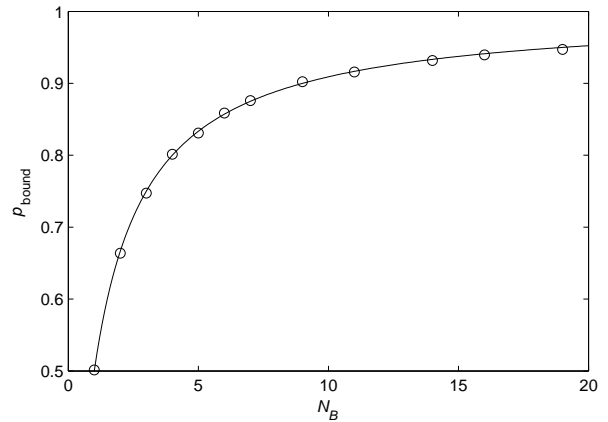} \caption{The probability $p_{\rm
bound}$ that the $A$ particle is bound to a $B$ particle as a function
of the total number of $B$ particles for the reaction scheme shown in
Eq.~\ref{eq:bimolreact}. The symbols indicate the simulation results,
while the solid line denotes the mean-field prediction
(Eq.~\ref{eq:pbound} with $g_{AB}(\sigma)=1$).  The radius of the
container is $R=200\sigma$ and the equilibrium constant is chosen such
that it is equal to the interaction volume $V$. The error bars in the
simulation results are smaller than the size of the
symbols. \label{fig:container}}
\end{figure}

The radial distribution function $g_{AB} (r)$ describes the spatial
correlations arising from the interactions between the
particles~\cite{HansenMcDonalds}. It is conceivable that in this
system the excluded volume interactions between the particles induce
spatial correlations. These correlations could affect the density of
$B$ particles that are in contact with the $A$ particle and thereby
the probability that the $A$ particle is bound to a $B$ particle. In
Eq.~\ref{eq:Keq}, the distribution function at contact, $g_{AB}
(\sigma)$, thus describes the effect of the spatial correlations on
the equilibrium constant. However, the concentrations that we consider
here are very low and, as a result, the spatial correlations are expected
to be small. Indeed, the simulations reveal that $g_{AB} (r) \approx 1$ for
all distances $r$.  If $g_{AB} (\sigma) = 1$, then
Eqs.~\ref{eq:pbound} and~\ref{eq:Keq} reduce to the well-known
mean-field results that can straightforwardly be obtained by solving
the macroscopic rate equations in steady-state. In
Fig.~\ref{fig:container}, we compare the simulation results to the
mean-field prediction for $p_{\rm bound}$. We find excellent agreement.

In conclusion, we have shown that our algorithm provides an accurate
 way of simulating an assembly of particles that can move by diffusion
 and react according to monomolecular and bimolecular reactions. As
 more complicated reactions, such as trimolecular reactions, can, in
 general, be decomposed into these elementary reactions, we are now in
 a position to simulate more complex systems.

\subsection{Gene expression}
In this section we present results for a model of gene expression. It
should be stressed that the model is highly simplified. The purpose
here is to show the power of our approach. Nevertheless, we find interesting
effects due to the spatial fluctuations of the components that could
be of relevance for more realistic systems.

The reaction network consists of the following reactions:
\begin{eqnarray}
\label{eq:geneI_1}
A+B &\overset{k_a}{\underset{k_d}{\rightleftarrows}}& C \\ 
C
&\underset{k_{\text{prod}}}{\rightarrow}& P + A + B
\label{eq:geneI_2}\\ P &\underset{k_{\text{dec}}}{\rightarrow}&
\varnothing
\label{eq:geneI_3}
\end{eqnarray}
In Eqs.~\ref{eq:geneI_1}--~\ref{eq:geneI_3}, $A$ represents a promoter
region on the DNA and $B$ a RNA polymerase molecule that moves by free
diffusion and can bind with a forward rate $k_a$ to the promoter site
to form the RNAp-DNA complex $C$. This complex can dissociate with a
rate constant $k_d$. Alternatively, it can produce a protein $P$ at a
production rate $k_\text{prod}$. Proteins degrade with a decay rate
$k_{\text{dec}}$. Note that, in this model, when a protein is produced
the RNAp dissociates from the DNA.

In the living cell, the concentration of free RNAp -- RNAp that is not
bound to the DNA -- is usually very low~\cite{Swain}. As a result,
spatial correlations are negligible (i.e., $g_{AB}(r) = 1$) and the
mean number of proteins, $\overline{ N_P}$, can be obtained from the
macroscopic rate equations. The result is:
\begin{equation}
\label{eq:np}
\overline{N_P} = K_1 K_2 \frac{N_{B}}{V + K_1 N_{B}}.
\end{equation}
Here $K_1=k_a/(k_d + k_\text{prod})$ and $K_2 =
k_\text{prod}/k_\text{dec}$ and $N_B$ is the total number of $B$
molecules.

In the simulations, we fix the promoter site, i.e. the $A$ particle,
in the center of a spherical container of radius $R$. The volume of
the container is $V=1 \mu {\rm m}^3$, which is comparable to that of
the {\em Escherichia coli} cell. The promoter site is surrounded by
$N_{B}=18$ RNAp molecules, corresponding to the concentration of free
RNAp of $30$ nM as found in the living cell~\cite{Swain}. The RNAp
molecules move with a diffusion constant $D=1 \mu {\rm m}^2 {\rm
s}^{-1}$, which is comparable to that of similarly sized
proteins~\cite{Elowitz99}. We assume that, at contact, the RNAp can
associate with the promoter site at a rate as determined by the
Maxwell-Boltzmann velocity distribution. This leads to $k_a = \pi
\sigma^2 \langle v_{AB} \rangle = 3\cdot10^9 {\rm M}^{-1} {\rm
s}^{-1}$, where $v_{AB}$ is the relative velocity of the particles $A$
and $B$. We note that this estimate is equal to the rate of collisions
between hard spheres in the low density limit
\cite{Szabo,HansenMcDonalds}.  We could arrive at an alternative
estimate for $k_a$ using the diffusion constant and a molecular
``jump'' distance $\lambda$. This would lead to $k_a = 4 \pi \sigma^2
D / \lambda$. Both estimates give similar results for the value of
$k_a$. The dissociation rate is chosen such that the equilibrium
constant $K=k_a/k_d$ equals the one reported in~\cite{Swain}, yielding
$k_d=21.5 s^{-1}$. We assume that the diameters of the promoter site
and the RNAp molecules are equal and given by $\sigma = 5 {\rm nm}$.

Here, we only simulate the promoter site and the RNAp molecules
explicitly in space. The proteins are assumed to be uniformly
distributed in space. Moreover, we reduce both the degradation and the
production of protein molecules to single-step Poisson
processes. These assumptions are unrealistic. Nevertheless, this model
allows us to demonstrate the power and the flexibility of our algorithm. In
particular, the production and decay reactions can simply be added to
our list of possible reactions, $\{R_\nu\}$ (see
section~\ref{sec:algorithm}). The next-reaction distribution function
for the production reaction is given by $q_{\rm prod} (t) = k_{\rm
prod} N_{C} \exp(-k_{\rm prod} N_{C} t) $, where $N_{C} = 0$ if the
RNAp is unbound and $N_{C} = 1$ if it is bound to the DNA, while the
propensity function for the degradation reaction is given by $q_{\rm
decay} (t) = k_{\rm decay} N_P \exp(-k_{\rm decay} N_P t)$. In this
way, the spatially-resolved GFRD scheme can naturally be combined with
kinetic Monte Carlo schemes, such as the Gillespie
algorithm~\cite{Gillespie76}, that are based upon the spatially
uniform chemical master equation.

\begin{figure}[b] \centering
\includegraphics[width=8cm]{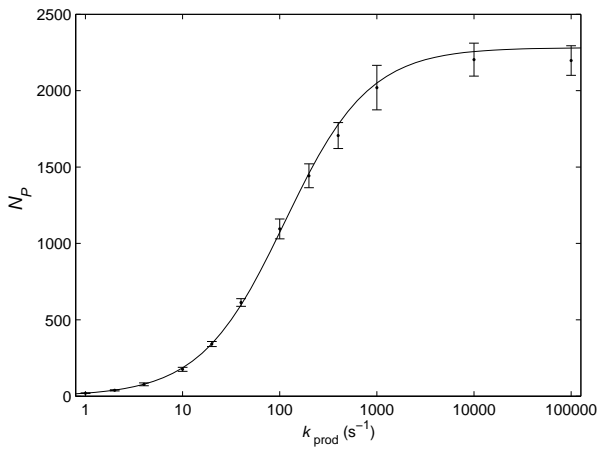} \caption{The mean protein number
$\overline{N_P}$ as a function of the protein production rate
$k_\text{prod}$ as obtained from the GFRD simulations for the reaction
scheme shown in Eqs.~\ref{eq:geneI_1} --~\ref{eq:geneI_3}. The solid line
denotes the mean-field prediction given by Eq.~\ref{eq:np}.}
\label{fig:np_vs_kprod}
\end{figure}

In figure \ref{fig:np_vs_kprod} we show the mean number of proteins
$\overline{N_P}$ as a function of the protein production rate
$k_\text{prod}$, while keeping the decay rate fixed at $k_{\rm decay}
= 0.04 s^{-1}$. As the concentration of the RNAp is low and spatial
correlations are expected to be negligible, the simulation results for
the {\em average} number of proteins, $\overline{N_P}$, should follow
the mean-field prediction of
Eq.~\ref{eq:np}. Fig.~\ref{fig:np_vs_kprod} shows that this is indeed
the case. However, in contrast to the mean-field analysis, the
GFRD simulations allow us to quantify the effect of the {\em spatial
fluctuations} of the RNAp molecules on the {\em noise} in the protein
synthesis.

We can quantify the magnitude of the noise in protein production by
 computing the following quantity~\cite{Swain}:
\begin{equation}
\eta^2_P = \frac{\overline{N_P^2 (t)}- \overline{N_P}^2}{\overline{N_P}^2}.
\end{equation} 
In the following
analysis we have changed the degradation rate such that the average
number of proteins, $\overline{N_P}$, is constant. This allows us to
focus on the effect of spatial fluctuations on the noise in protein
production -- we thus eliminate the fairly trivial changes in
the noise due to changes in the average number of proteins.

Since we are interested in the importance of spatial fluctuations in
gene expression, it is natural to compare the GFRD results to those
obtained using the chemical master equation. The latter approach does
take into account the discrete nature of the reactants and the
stochastic character of their interactions, but it treats the spatial
fluctuations in a mean-field manner: at each instant, it is
implicitly assumed that the particles are uniformly distributed in
space. This approach is justified if there are many non-reactive
collisions to stir the system in between the reactive
collisions. However, the RNAp is present in low copy numbers, and,
upon contact, it rapidly associates with the promoter site on the
DNA. As a consequence, this reaction is diffusion-limited. This could
have important implications for the noise in gene expression.

Using the techniques described in~\cite{VanKampen}, we can
analytically obtain the noise from the chemical master
equation. It is given by
\begin{equation}
\eta_P^2=\frac{1}{\overline{N_P}}-\frac{k_\text{prod} k_a
N_{B}}{k_\text{prod} k_a N_{B} + \overline{N_P} (k_a N_{B} + k_d +k_\text{prod})^2}.
\label{eq:noise}
\end{equation}
The first term on the right describes the result that would have been
obtained if gene expression were a simple linear birth-and-death
process. The second term reflects the fact that in order to produce a
protein, it is necessary, albeit not sufficient, for a RNAp molecule
to bind to the promoter site. This term, and thus the noise in gene
expression, goes through a minimum at $k_{\rm prod} = k_a N_{B} + k_d$
and vanishes for both small and large $k_{\rm prod}$. In these
regimes, gene expression reduces to a simple linear birth-and-death
process. In the limit of small $k_{\rm prod}$, the production of the
protein is the rate limiting step. The RNAp molecule will associate to
and dissociate from the promoter site a large number of times, before
it actually induces gene expression. The former process is thus in
equilibrium on the time scale of gene expression. Hence, the birth
term is given by $k_{\rm birth} = p_{\rm bound} k_{\rm prod}$, with
$p_{\rm bound}$ being the probability that a RNAp molecule is bound to
the promoter (see Eq.~\ref{eq:pbound}); the death term is
given by $k_{\rm death} = k_{\rm decay}$. In the limit of large
$k_{\rm prod}$, the binding of a RNAp molecule to the promoter site
is the rate limiting step: as soon as a RNAp molecule is bound to the
promoter, a protein will be produced. This means that the birth term
is given by $k_{\rm birth} = k_a (1-p_{\rm bound}) [B]$; the
death term is again $k_{\rm death} = k_{\rm decay}$. For a linear
birth-and-death process, the noise is determined by the average number
of proteins, $\eta_P = 1/\sqrt{\overline {N_P}}$~\cite{VanKampen}. As we
have set the decay rate $k_{\rm decay}$ such that $\overline{N_P}$ is constant,
the noise in gene expression must be the same in the limiting regimes
of small and large $k_{\rm prod}$, in which gene expression reduces to
a birth-and-death process.

\begin{figure}[b] \centering
\includegraphics[width=8cm]{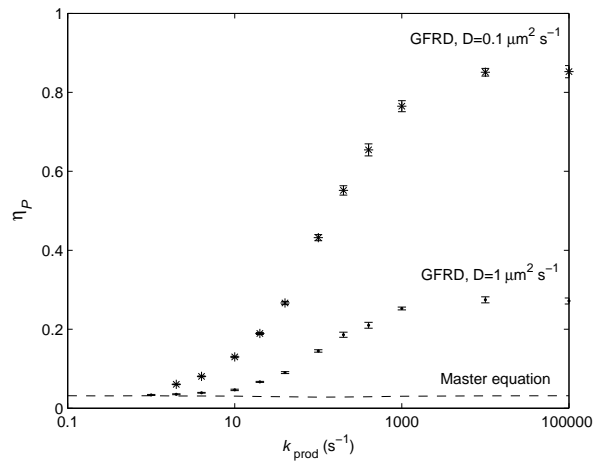} \caption{The noise in protein
level $\eta_P$ as a function of protein production rate
$k_\text{prod}$ for the reaction scheme shown in Eqs.~\ref{eq:geneI_1}
--~\ref{eq:geneI_3}. Compared are the results obtained by GFRD with a
diffusion constant of $D=1\mu {\rm m}^2 {\rm s}^{-1}$ ($\cdot$) and
$D=0.1\mu {\rm m}^2 {\rm s}^{-1}$ ($*$) and the result using the
chemical master equation (dashed line). The mean number of proteins
was held constant at $\overline{N_P}=1000$ by changing the degradation
rate $k_{\rm decay}$ of the protein.} \label{fig:GvsGFBD}
\end{figure}

Figure~\ref{fig:GvsGFBD} shows the noise in
the protein concentration as a function of the synthesis rate for
$\overline{N_P}=1000$. The GFRD results are compared to
those obtained using the chemical master equation. It is seen that for
small $k_\text{prod}$ both approaches yield identical results. In this
regime, protein synthesis is the rate-limiting step. Indeed, on the
time scale of gene expression the RNAp molecules have sufficient time
to become well mixed in the cell. As a result, the effects of
diffusion are negligible and the noise reduces to the expected value
for a linear birth-and-death process.

However, for $k_{\rm prod} \gtrsim 1 s^{-1}$, spatial fluctuations can
have a drastic effect on the noise in gene expression. In this
regime, the noise of the spatially-resolved model is larger than that
of the ``well-stirred reactor'' model. In fact, Fig.~\ref{fig:GvsGFBD}
shows that it grows fairly rapidly with increasing
$k_\text{prod}$. The increase in noise is due to a very broad
distribution of arrival times of RNAp molecules at the promoter site,
much broader than the corresponding Poisson distribution for the
system without spatial fluctuations. It is also seen that for very
large production rates, the noise ultimately reaches a plateau
value. At these high values of $k_{\rm prod}$ the promoter site
becomes an ``absorbing'' boundary for the RNAp molecules.
Fig.~\ref{fig:GvsGFBD} also reveals that the height of the plateau
increases as the diffusion constant $D$ becomes smaller. This is not
surprising, because the importance of spatial fluctuations is expected
to be larger for smaller diffusion constants. However, it does clearly
show that in order to determine the significance of spatial
fluctuations in gene expression, it is of interest to establish the
value of the diffusion constant of the RNA polymerase experimentally.

This model of gene expression is obviously highly simplified -- our
aim here is to present a new particle-based technique to simulate
biochemical networks. Nevertheless, the results show that spatial
fluctuations are potentially important in gene expression. Moreover,
GFRD will now make it possible to study systematically how significant
the effects of spatial fluctuations are on the noise in gene expression
using more refined models.

\subsection{Performance}
\label{sec:performance}
The essence of the GFRD scheme is to exploit the analytical solution
of the Smoluchoswki equation for a pair of interacting particles to
set up an event-driven algorithm. This allows GFRD to make large jumps
in time when the particles are far apart from each other. Clearly, the
performance of the algorithm depends on the density of the system: the
further the particles are apart, the larger the time
step that can be used and the better GFRD will perform in comparison
to brute-force Brownian Dynamics schemes (see
section~\ref{sec:tech_intro}). It is thus of interest to compare the
distribution of propagation times in GFRD to the time step used in a
brute-force Brownian Dynamics scheme as a function of density.

\begin{figure}[t] \centering
\includegraphics[width=8cm]{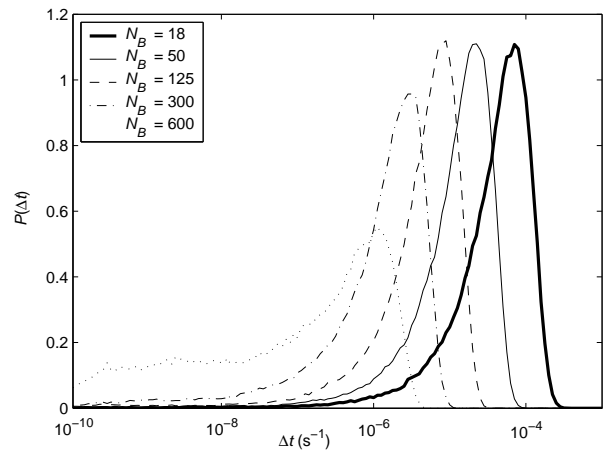} \caption{The distribution of
propagation times $\Delta t$ for a system consisting of a single
particle $A$ in the center of a spherical box of volume $V=1 \mu {\rm
m}^3$, surrounded by $N_B$ particles $B$; the particles $A$ and $B$
can react according to a bimolecular reaction scheme (see
Eq.~\ref{eq:bimolreact}). The distributions are shown for $N_B=18$
(bold solid line), $N_B = 50$ (solid line) $N_B = 125$ (dashed line),
$N_B = 300$ (dashed-dotted line) and $N_B = 600$ (dotted line),
corresponding to a concentration of $[B]=30$ nM, $[B] = 83$ nM, $[B] =
210$ nM, $[B] = 500$ nM, and $[B] = 1000$ nM, respectively. In the
GFRD simulations, we use a lower cut-off for the propagation time,
$2.5\cdot10^{-10} {\rm s}$, which corresponds to the time step used in a
brute-force Brownian Dynamics simulation.}
\label{fig:t_prop}
\end{figure}

In figure \ref{fig:t_prop} we show the distribution of propagation
times $\Delta t$ for the bimolecular reaction described in section
\ref{sec:bimol} as a function of density.  For $N_B=18$ ($[B] = 30$
nM), the value used in the above models of gene expression, the
distribution has a maximum at $\Delta t = 1\cdot10^{-4} {\rm s}$. The
propagation times become smaller if the density increases. For $[B] =
1 \mu {\rm M}$, the peak in the distribution shifts to $\Delta t =
1\cdot10^{-6} {\rm s}$.

In biochemical networks, the concentrations of the components can be
very low. In gene networks, for example, the concentrations of the
gene regulatory proteins are often in the nM regime. In signal
transduction networks, the concentrations of the components may also
be fairly low, i.e. in the $\mu {\rm M}$ range.  The analysis
presented here suggests that with GFRD it should be possible to reach
time steps of at least $10^{-6} - 10^{-4} {\rm s}$ for such
networks. In contrast, in a brute-force Brownian Dynamics simulation,
we cannot use a time step larger than $10^{-10} - 10^{-9} {\rm s}$
($10^{-5} - 10^{-4} \sigma^2/D$) in order to preserve the correct
distribution (as, for instance, defined by the requirement that the
analytical solution for the Green's function $p_{\rm rev}({\bf
r},t|{\bf r}_0,t_0)$ as shown in Fig.~\ref{fig:Gf} can be accurately
reproduced). We thus believe that under biologically relevant
conditions, GFRD can be three to six orders of magnitude faster than
conventional particle-based schemes for simulating biochemical
networks in time and space.

\section{Conclusions}
We have developed a new technique, called Green's Function Reaction
Dynamics, to simulate biochemical networks at the particle level and
in both time and space.  The main idea of the technique is to choose
a maximum time step such that only single particles or pairs of particles
have to be considered. For these particles, the Smoluchowski equation
can be solved analytically using Green's functions. The analytical
solution can then be used to set up an event-driven algorithm, quite
analogous to the kinetic Monte Carlo schemes as originally developed
by Bortz, Kalos and Lebowitz~\cite{Bortz75} to simulate Ising spin
systems and by Gillespie to numerically solve the chemical master
equation~\cite{Gillespie76}. We would like to stress,
however, that in contrast to the widely used ``Gillespie'' algorithm,
our technique makes it possible to simulate biochemical networks at
the particle level and in both time and {\em space}.

The analysis presented in section~\ref{sec:performance} shows that
GFRD is highly efficient. This should make it possible to simulate
biochemical networks at much larger length and time scales than
hitherto possible. In addition, we believe that the scheme has the
potential to be even more efficient. In the current scheme, we use a
global maximum time step that pertains to all particles in the
system. It seems natural, however, to assign to each particle an
individual maximum time step.  In such a scheme, each particle would
have its own individual clock. This approach would make it possible to
devote most computational effort to those particles that interact
frequently; the particles that are initially far from other
particles are only updated once the time has come when they have a
reasonable chance to interact. A second possible improvement would be
to exploit the low concentration of the components in another way. In
the current scheme, we explicitly take into account excluded volume
interactions. In fact, this often imposes a limit on the maximum time
step. If the concentrations are low, however, we would expect the
excluded volume effects to be negligible for the behavior of the
network. In future work, we will show how these observations can be
incorporated into the algorithm to even further enhance the
performance of Green's Function Reaction Dynamics.
 
\section*{Acknowledgments} We would like to thank Marco Morelli,
Siebe van Albada and Daan Frenkel for
useful discussions and suggestions and Rosalind Allen and Bela Mulder
for a critical reading of the manuscript. The work is part of the
research program of the ``Stichting voor Fundamenteel Onderzoek der
Materie (FOM)", which is financially supported by the ``Nederlandse
organisatie voor Wetenschappelijk Onderzoek (NWO)".\\

\section*{Appendix A: Solution of Smoluchowski equation}
Here we consider a pair of particles $A$ and $B$ that move by free
diffusion, but, upon contact, can react with a rate constant $k_a$.
The Green's function $p_2^{\bf r}({\bf r},t|{\bf r}_0,t_0)$ for this
pair can be obtained by exploiting the analogy between the diffusion
of particles and the conduction of heat in solids. The corresponding
Green's function for the conduction of heat in solids is derived
in~\cite{Carslaw}. The Green's function $p_2^{\bf r}({\bf r},t|{\bf
r}_0,t_0)$ is
\begin{widetext}
\begin{equation}
\label{eq:pGr}
p_2^{\bf r}(r,\theta,\phi,t|r_0)=\frac{1}{4 \pi \sqrt{r r_0}}\sum^{\infty}_{n=0}(2n+1) P_n(\cos \theta)
						\int^{\infty}_{0}e^{-D u^2 t} F_{n+1/2}(u r) F_{n+1/2}( u r_0) u du, 
\end{equation} 
where
\begin{equation}
F_\nu(u r)=\frac{(2 \sigma k_a +1)[J_\nu(u r) Y_\nu(u \sigma)-Y_\nu(u
r) J_\nu(u \sigma)] -2 u \sigma [J_\nu(u r) Y'_\nu(u \sigma)-Y_\nu(u
r) J'_\nu(u \sigma)]} { ( [ (2 \sigma k_a +1)J_\nu(u \sigma)-2u\sigma
J'_\nu(u \sigma)]^2 + [ (2 \sigma k_a +1)Y_\nu(u \sigma)-2u\sigma
Y'_\nu(u \sigma)]^2 )^{1/2} },
\end{equation}
\end{widetext}
and where $P_n$ is the $n$th Legendre polynomial, $J_n$ and $Y_n$ are
the $n$th Bessel function of the first and the second kind, $D = D_A +
D_B$ is the total diffusion constant of the two particles $A$ and $B$
and $\sigma = (d_A + d_B)/2$, where $d_A$ and $d_B$ are the diameters of
the particles $A$ and $B$, respectively; here, and below, we take $t_0 =
0$. The Green's function can be
expressed in a more compact notation by
\begin{equation} 
p_2^{\bf r}(r,\theta,\phi,t|r_0)=\sum^{\infty}_{n=0}C_n P_n(\cos\theta)R_n(r, t).
\end{equation}
The probability $f(r|r_0, t)$ of finding the particles separated by a
distance between $r$ and $r+dr$ at time $t$ is given by 
\begin{equation} \label{eq:f}
f(r|r_0, t) = 2 \pi \sum_{n=0}^\infty C_n Q_n(\pi) r^2 R_n(r,t),
\end{equation} 
with
\begin{equation}
Q_n(\theta) = \int^\theta_0 \sin \theta P_n(\cos\theta)d\theta.
\end{equation}
The conditional probability $g(\theta|r, r_0, t)$ that two particles are at
an angle between $\theta$ and $\theta+d\theta$ with respect to the
original direction ${\bf r}_0={\bf r}_B-{\bf r}_A$, {\em given} that
they are separated by a distance $r$ at time $t$, is 
\begin{equation} \label{eq:g}
g(\theta|r, r_0, t) = 2 \pi \sum_{n=0}^\infty C_n Q'_n(\theta) r^2 R_n(r,t).
\end{equation}
The survival probability $S(t)$ is given by
\begin{equation}
S(t)=\int_\sigma^\infty f(r|r_0, t) dr.
\end{equation}
The above integral is complicated but it follows from the properties of the diffusion equation that it must be identical to the familiar survival probability for the spherically symmetric case~\cite{Rice, Kim}.

The above distribution functions suggest a straightforward procedure
for drawing a new position ${\bf r}$ from the Green's function
$p_2^{\bf r}({\bf
r},t|{\bf r}_0)$. We pretabulate $Q_n(\theta)$ and $R_n(r,r_0,t)$ up to
a certain order $N$. From this we construct the probability
distribution $f(r|r_0, t)$ and we draw a new distance $r$ from this
distribution. Next, we draw $\theta$ from the distribution
$g(\theta|r, r_0, t)$ and finally we choose $\phi$ uniformly distributed
between $0$ and $2\pi$.

This procedure works well for large $t$. For small $t$, however,
the above procedure becomes rather cumbersome as the number
of terms $N$ that needs to be included in order for the summations in
Eqs.~\ref{eq:f} and~\ref{eq:g} to converge, becomes very large. The
reason is that $p_2^{\bf r}({\bf r},t|{\bf r}_0)$ becomes a sharply peaked function
around ${\bf r}_0$ for small $t$. However, for small $t$, the
probability that the two particles will interact with each other, is
relatively small. In other words, for small $t$ the full solution
$p_2^{\bf r}({\bf r}, t|{\bf r}_0)$, is dominated by free diffusion. We can exploit
this observation in order to reduce $N$ by writing the Greens's
function as $p_2^{\bf r}({\bf r}, t|{\bf r}_0) = p_{\rm free}({\bf r}, t|{\bf r}_0) + p_{\rm
corr}({\bf r}, t|{\bf r}_0)$, where $p_{\rm free}$ is the solution for free
diffusion and $p_{\rm corr}$ is a correction term that takes into
account the reacting boundary at $r=\sigma$. The free diffusion term
can easily be computed from
\begin{widetext}
\begin{equation}
p_{\rm free} (r,\theta,\phi,t| r_0) = 
 \frac{1}{\left[4 \pi D
t\right]^{3/2}} \exp\left[-\frac{r^2 + r_0^2 - 2 r r_0 \cos \theta}{4 D
t}\right].
\end{equation} 
Using the fact, that $p_{\rm free}$ can also be written as
\begin{equation}
p_{\rm free}(r,\theta,\phi,t|r_0)=\frac{1}{4 \pi \sqrt{r r_0}}\sum^{\infty}_{n=0}(2n+1) P_n(\cos \theta)
						\int^{\infty}_{0}e^{-D u^2 t} J_{n+1/2}(u r) J_{n+1/2}( u r_0) u du, 
\end{equation} 
we find by comparison with Eq.~\ref{eq:pGr} that $p_{\rm corr}$ can be expressed as
\begin{equation}
p_{\rm corr}(r,\theta,\phi,t|r_0)=-\frac{1}{4 \pi \sqrt{r r_0}}\sum^{\infty}_{n=0}(2n+1) P_n(\cos \theta)
						\int^{\infty}_{0}e^{-D u^2 t} \frac{R_1}{R_1^2+R_2^2}( R_1 F_1 + R_2 F_2) u du, 
\end{equation} 
\end{widetext}
where
\begin{eqnarray}
R_1 &=& (2 \sigma k_a \!+\! 1 ) J_{n\!+\!\frac{1}{2}}(u \sigma) - 
	2 u \sigma J'_{n\!+\!\frac{1}{2}}(u \sigma), \\
R_2 &=& (2 \sigma k_a \!+\! 1 ) Y_{n\!+\!\frac{1}{2}}(u \sigma) - 
	2 u \sigma Y'_{n\!+\!\frac{1}{2}}(u \sigma), \\
F_1 & = & J_{n\!+\!\frac{1}{2}}(u r) J_{n\!+\!\frac{1}{2}}(u r_0) - 
	Y_{n\!+\!\frac{1}{2}}( u r) Y_{n\!+\!\frac{1}{2}} (u r_0),\\
F_2 & = & J_{n\!+\!\frac{1}{2}}(u r) Y_{n\!+\!\frac{1}{2}}(u r_0) + 
	J_{n\!+\!\frac{1}{2}}( u r_0) Y_{n\!+\!\frac{1}{2}} (u r).
\end{eqnarray}
As the correction term
$p_{\rm corr}$ is usually rather small for small $t$, the number of
terms $N$ that needs to be included in  order to compute the functions
$f(r,|r_0,t)$ and $g(\theta|r,r_0,t)$,  is strongly reduced.

\section*{Appendix B: Proof of the algorithm}
Let $P(t,\mu)dt$ be the probability that the next reaction will occur
in the time interval between $t$ and $t+dt$ {\em and} will be reaction
$R_\mu$. As described in section~\ref{sec:overview}, the time step is
chosen such that the reactions occur {\em independently} from each
other. The probability $P(t,\mu)dt$ is therefore given by
\begin{equation}
\label{eq:Ptmu}
	P(t,\mu) dt = q_\mu (t) dt \prod^M_{\nu=1 \atop \nu\neq\mu} S_\nu (t).
\end{equation}
The algorithm features a
maximum time step. It is thus conceivable that not a single reaction
occurs within a time step. The probability $Q (\Delta t_{\rm max})$
that no reaction occurs within a time step of size $\Delta t_{\rm
max}$ is given by
\begin{equation}
\label{eq:Q}
Q(\Delta t_{\rm max}) = \prod^M_{\nu=1} S_\nu (\Delta t_{\rm max}).
\end{equation}

It can easily be shown that the procedure outlined in
section~\ref{sec:algorithm} is consistent with
Eqs.~\ref{eq:Ptmu} and~\ref{eq:Q}. Let $\check{Q}(\Delta t_{\rm max})$
be the probability that the procedure described above does not yield a reaction
within the time step of size $\Delta t_{\rm max}$. It is given by
\begin{eqnarray}
\check{Q} (\Delta t_{\rm max}) &=& \mbox{Prob } (t_\nu > \Delta t_{\rm
max} \mbox{ for all }
\nu)\\
&=&\prod_{\nu=1}^{M} \mbox{Prob } (\xi_\nu > (1-S_\nu (\Delta t_{\rm max})))\\
&=&\prod_{\nu=1}^{M} S_\nu(\Delta t_{\rm max})\\
&=&Q (\Delta t_{\rm max}).
\end{eqnarray} 
Similarly, let $\check{P}(t,\mu) dt$ be the probability that the above
described procedure yields, at time $t$, reaction $R_\mu$ as the next
reaction . It is given by
\begin{eqnarray}
\check{P}(t,\mu)dt &=&\mbox{Prob } (t<t_\mu <t+dt) \nonumber \\
& & {} \times \mbox{Prob } (t_\nu >
t_\mu \mbox{ for all } \nu \neq \mu)\\
& = & q_\mu (t) dt \prod_{\nu=1 \atop \nu \neq \mu}^M \mbox{Prob } (\xi_\nu > (1-S_\nu (t_\nu)))\\
&=& q_\mu (t) dt \prod_{\nu=1 \atop \nu \neq \mu}^M S_\nu (t_\nu)\\
&=& P(t,\mu) dt.
\end{eqnarray}


\begin{thebibliography}{99}
\bibitem{McAdams97} McAdams and A. Arkin, Proc. Natl. Acad. Sci. USA,
{\bf 94}, 814 (1997).
\bibitem{Elowitz00} M. B. Elowitz and S. Leibler, Nature (London)
{\bf 403}, 335 (2000); N. Barkai and S. Leibler, Nature (London), {\bf
403}, 267 (2000).
\bibitem{VanKampen} N. G. van Kampen, {\em Stochastic Processes in Physics and Chemistry}, North-Holland, Amsterdam (1992).
\bibitem{Togashi} Y. Togashi and K. Kaneko, Phys. Rev. Lett {\bf 86}, 2459 (2001).
\bibitem{Shnerb} N. M. Shnerb, Y. Louzoun, E. Bettelheim and
S. Solomon, Proc. Natl. Acad. Sci. {\bf 97}, 10322 (2000).
\bibitem{Gillespie76} D. T. Gillespie, J. Comput. Phys. {\bf 22}, 403
(1976); D. T. Gillespie, J. Phys. Chem. {\bf 81}, 2340
(1977).
\bibitem{Bray} C. J. Morton-Firth and D. Bray, J. Theor. Biol. {\bf 192},  117 (1998).
\bibitem{Albertsbook} B. Alberts, D. Bray, J. Lewis, M. Raff,
J. Lewis, M. Raff, K. Roberts, and J. D. Watson, {\em Molecular
Biology of the Cell}, Garland Publishing, New York, (1994).
\bibitem{Smoluchowski17} M. Smoluchowski, Z. Phys. Chem. {\bf 92}, 129
(1917); S. Chandrasekhar, Rev. Mod. Phys. {\bf 15}, 1 (1943).
\bibitem{Kim} H. Kim and K. J. Shin, Phys. Rev. Lett. {\bf 82}, 1578 (1999).
\bibitem{Carslaw} H. S. Carslaw and J. C. Jaeger, {\em Conduction of
Heat in Solids}, Oxford University Press, New York(1959).
\bibitem{Agmon90} N. Agmon, A. Szabo, J. Chem. Phys. {\bf 92}, 5270
(1990).
\bibitem{HansenMcDonalds} J.-P. Hansen \& I. R. McDonald, {\em Theory
of simple liquids}, 2nd edn, Academic Press, San Diego (1986).
\bibitem{Elowitz99} M. B. Elowitz, M. G. Surette, P.-E. Wolf,
J. B. Stock, and S. Leibler, J. Bacteriol. {\bf 181}, 197 (1999).
\bibitem{Swain} P. S. Swain, M. B. Elowitz and E. D. Siggia, Proc. Natl. Acad. Sci. {\bf 99}, 12795 (2002).
\bibitem{Szabo} H. X. Zhou and A. Szabo, J. Chem. Phys. {\bf 95}, 5948 (1991).   
\bibitem{Rice} S. A. Rice, {\em Diffusion Limited Reactions}, Elsevier Science Publishing, New York (1985).

\bibitem{Ptashne02} M. Ptashne \& A. Gann {\em Genes and signals}.
Cold Spring Harbor Laboratory Press, New York (2002).
\bibitem{Bortz75} A. B. Bortz, M. H. Kalos, and J. L. Lebowitz,
J. Comp. Phys. {\bf 17}, 10 (1975).
\end{thebibliography}
\end{document}